# Relativistic Jets from Black Holes: A Unified Picture

C Sivaram and Kenath Arun

Indian Institute of Astrophysics, Bangalore

**Abstract:** The current understanding of the formation of powerful bi-directional jets in systems such as radio galaxies and quasars is that the process involves a supermassive black hole that is being fed with magnetized gas through an orbiting accretion disc. In this paper we discuss the dynamics of the jet powered by rotating black holes, in the presence of a magnetic field, including the scaling of the jet length and their typical time scales. We consider a unified picture covering all phenomena involving jets and rotating black holes ranging from gamma ray bursts to extragalactic jets and discuss the relevant scaling laws. We have also discussed the acceleration of the particles in jets and consequent synchrotron and inverse Compton radiations. Accelerated protons from jets as possible sources of high energy cosmic rays are also discussed.

## 1. Introduction

Objects as diverse as X-ray binaries, radio galaxies, quasars, and even the galactic centre, are powered by the gravitational energy released when surrounding gas is accreted by the black hole at their cores. The combination of the strong gravity of the black hole, the rotation in the in falling matter, and the magnetic field are believed to be the key ingredients to jet creation. Apart from copious radiation, one of the manifestations of this accretion energy release is the production of jets, collimated beams of matter that are expelled from the innermost regions of accretion discs. These jets shine particularly brightly at radio frequencies.[1, 2]

An accretion disc is matter that is drawn to the black hole. In rotating black holes, the matter forms a disc due to the mechanical forces present. In a Schwarzschild black hole, the matter would be drawn in equally from all directions, and thus would form an omni-directional accretion cloud rather than a disc.[3]

Jets form in Kerr black holes that have an accretion disc. The matter is funnelled into a disc-shaped torus by the black hole's spin and surrounding magnetic fields, but in the very narrow regions over the black hole's poles, matter can be energized to extremely high temperatures and speeds, escaping the vicinity of the black hole in the form of high-speed jets.[3, 5]

## 2. Dynamics of the Jet

For a Kerr black hole, the horizon is given by:[6]

$$r = m \pm \sqrt{m^2 - a^2} \qquad \ldots (1)$$

Where, $m$ is the geometric mass and $a$ is the geometric angular momentum. From the condition that $r$ should be real, the limiting case is given by, $m = a$. That is:

$$\frac{GM}{c^2} = \frac{J_{Max}}{Mc^2} \qquad \ldots (2)$$

From this, the maximum angular momentum is given by,

$$J_{Max} = \frac{M^2 G}{c} \qquad \ldots (3)$$

From the classical expression for the angular momentum associated with a jet of length $l$, assuming the particles to be travelling at near speed of light, the expression becomes $J = mcl$

Considering a conical jet with base radius $r$ and density $\rho$, the mass of the jet is given by, $m = \frac{1}{3}\pi r^2 l \rho$

Then the angular momentum becomes:

$$J = \frac{1}{3}\pi l^2 r^2 c \rho \qquad \ldots (4)$$

From the geometry of the jet, we can relate the length of the jet to the radius $r$ as $r = l \tan 5^0$. Here we have assumed the small opening angle of the jet to be 5°. Assuming the number density of the jet to be of the order of $10^3 \, cm^{-3}$, which is consistent with observations, the length of the jet is given by

$$l = \left( \frac{3GM^2}{\pi \rho c^2 (\tan 5)^2} \right)^{1/4} \qquad \ldots (5)$$

For the billion solar mass black holes, it works out to $l = 6 \times 10^{19} \, m = 2 \, kpc$

The scaling of the length of the jet with the mass of the central black hole is given by:[7]

$$l = \left( \frac{3G}{\pi \rho c^2 (\tan 5)^2} \right)^{1/4} M^{1/2} \qquad \ldots (6)$$

$$l \approx 0.5\sqrt{M}$$

| M (in solar mass) | $l$ (parsec) |
|---|---|
| $10^2$ | 2 |
| $10^4$ | 20 |
| $10^6$ | 200 |
| $10^8$ | 2000 |

Table 1: Scaling of the length of the jet with the mass of the black hole

The behaviour of the length of the jet as a function of the mass of the black hole is shown in the following graph based on equation (6).

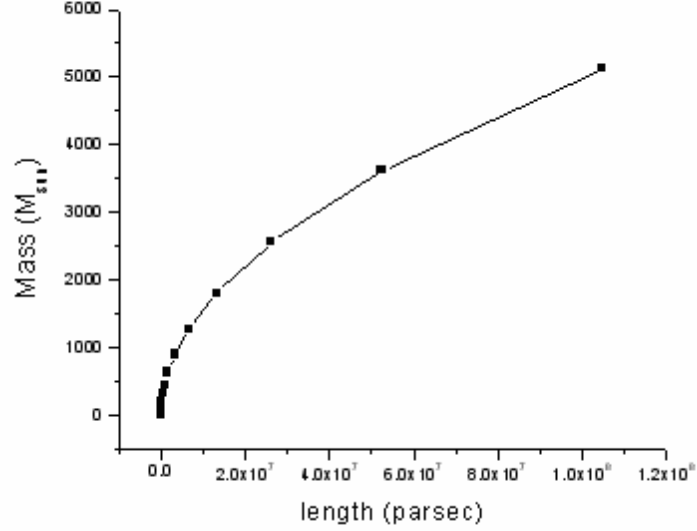

Fig 1: Length of the jet with BH mass

In the above discussion we have considered the variation of the length of the jet with the mass of the black hole. For a given black hole, the length of the jet depends on the density of the particles emitted out along the jet. From equation (5) we have

$$l = \left(\frac{3GM^2}{\pi c^2 (\tan 5)^2}\right)^{1/4} \frac{1}{\rho^{1/4}}$$

For a 30 solar mass black hole, with number density of $n = 10^{20} \, cm^{-3}$, the length of the jet is of the order of, $l = 10^{10} \, m$

| Number density, n(m⁻³) | Length of the jet, l |
|---|---|
| $10^3$ | 1kpc |
| $10^5$ | 300pc |
| $10^7$ | 100pc |
| $10^{20}$ | 0.01pc |

Table 2: Variation of jet length with number density (for billion solar mass black holes)

## 3. Jets in the presence of Magnetic Fields

The above discussion did not consider the presence of magnetic fields. We could also include the effects of magnetic (and electric) fields. We give a general picture in what follows.[8, 9]

The physical arguments can also encompass jets produced by short lived energetic events like gamma ray bursts (GRBs). For such events, the duration of the event $\Delta t \sim 1s$ and the gamma factor $\Gamma = (1-\beta^2)^{-1/2} \approx 300$, $\left(\beta = \dfrac{v}{c}\right)$.

As the objects involved in such highly energetic events are expected to be compact objects (for instance in short duration GRB, we have collisions (mergers) between neutron stars), where the surface magnetic fields (at a radius of $\sim 10^6 cm$) are considered very high $\sim 10^{16} G$.

Assuming flux $(BR^2)$ conservation, the field at a distance of $10^{10} cm$ is $\sim 10^8 G$.

With the presence of a magnetic field the electromagnetic power associated with the jet is given by:[9]

$$P = B^2 R^2 c \Delta t \Gamma^2 \qquad \ldots (7)$$

With the above values the power associated with the jet is of the order of $10^{51} ergs/s$. The maximum angular momentum for a ten solar mass black hole central engine (possible progenitor for a GRB) is given by equation (3) as:

$$J_{Max} = \dfrac{M^2 G}{c} \approx 10^{50} ergs.s$$

The time of rotation for the black hole is given by:[3, 5]

$$t_{Rot} = \dfrac{GM}{c^3} \qquad \ldots (8)$$

For the ten solar mass black hole this works out to be $t_{Rot} \sim 10^{-4} s$.

If the source of the energy is the rotation of the black hole, then the total energy radiated in this time scale is given by:[8, 10, 11]

$$E = \dfrac{J_{Max}}{t_{Rot}} \approx 10^{54} ergs \qquad \ldots (9)$$

This shows that the above jet can be powered with the rotational energy of the black hole central engine for several hundred seconds, which is consistent with observation.

We can also extend the above analysis and arguments to the case of the galactic jets. Here the magnetic field is much smaller, but the jet size is correspondingly much larger. The power associated with these jets is given by the same expression as equation (7), but magnetic field is $\sim 10^{-6} G$ and the extent $\sim 10^{25} cm$ and the gamma factor ~10.
The power thus works out to be of the order of $P \sim 3 \times 10^{50} ergs/s$.

The rotation of a billion solar mass black hole at the AGN centre gives a maximum angular momentum of:[3, 4, 5]

$$J = \frac{M^2 G}{c} \approx 10^{67} ergs.s \qquad \ldots (10)$$

The time of rotation scales with the mass and is of the order of $5 \times 10^3 s$ (as given by equation (8)). The total energy then becomes:

$$E = \frac{J}{t_{Rot}} \approx 10^{63} ergs \qquad \ldots (11)$$

This gives a time scale for the Mega-parsec jet of $\sim \frac{E}{P} \approx 10^{13} s \sim 10^6$ years, again consistent with observations.

These results are in excellent agreement with recent observations form MS0735.6+7421, where the energies and time scales are same as that obtained in this discussion.[12]

**4. Acceleration of the Particles in the Jet**

The potential across which the particles of the jet are accelerated can be determined as follows. Since the magnetic flux $(BR^2)$ is conserved, the magnetic field scales with the distance as $1/R^2$. The electric field is given by: $|E_0| = \frac{v}{c} B_0$ and since $v \approx c$, $E_0 = B_0$ and has the same $1/R^2$ scale dependence. At the surface corresponding to $R_0 = 10^6 cm$,

the magnetic field $B_0 \sim 10^{16} G$, which implies from the above argument that $E_0 \approx 10^{16} V/cm$.

Therefore the potential can be written as:

$$V = \int \frac{E_0 R_0^3}{x^3} dx \qquad \ldots (12)$$

For a distance at $10^{10} cm$ from the surface, the potential becomes:

$$V = \int_{10^6}^{10^{10}} \frac{E_0 R_0^3}{x^3} dx = 2 \times 10^{16} V \qquad \ldots (13)$$

Thus the energy associated with the electrons in the jet is of the order of $2 \times 10^{16} eV$.

The momentum of the particles (of rest mass $m_0$) in the jet is given by:

$$p = \gamma m_0 v \qquad \ldots (14)$$

Where the Lorentz factor for the particle is given by $\gamma = \frac{E}{m_0 c^2}$

The relativistic energy-momentum relation gives:

$$E^2 = p^2 + m_0^2 c^4 \qquad \ldots (15)$$

The Lorentz factor can then be written as:

$$\gamma^2 = \left(\frac{E}{m_0 c^2}\right)^2 = \frac{p^2 + m_0^2 c^4}{m_0^2 c^4} = 1 + \frac{p^2}{m_0^2 c^4} \qquad \ldots (16)$$

If the force acting on the particles of the jet is $F$, then:

$$p = \gamma m_0 v = Ft \qquad \ldots (17)$$

The velocity of the particles in the jet can then be written as:

$$v = \frac{Ft}{m_0 \gamma} = \frac{Ft}{m_0}\left(1 + \frac{F^2 t^2}{m_0^2 c^4}\right)^{-\frac{1}{2}} \qquad \ldots (18)$$

For small time t, the second term in the bracket vanishes giving:

$$v = \frac{ds}{dt} = \frac{Ft}{m_0} \qquad \ldots (19)$$

Solving the above equation we get the parabolic equation:

$$s = \frac{1}{2}\frac{F}{m_0}t^2 = \frac{1}{2}at^2 \qquad \ldots (20)$$

Where, $a = \dfrac{F}{m_0} = \dfrac{eE_0}{m_0}$ is the acceleration of the particles in the jet.

For an electron subjected to an electric field $E_0 \approx 10^{16} V/cm$, the acceleration is of the order of $10^{33} cm/s^2$.

The general solution for equation (17) is given by:[2, 6]

$$s = \int ds = \dfrac{F}{m_0}\int t\left(1 + \dfrac{F^2 t^2}{m_0^2 c^4}\right)^{-1/2} dt = \dfrac{m_0 c^4}{F}\left[\left(1 + \left(\dfrac{Ft}{m_0 c^2}\right)^2\right)^{1/2} - 1\right] \qquad \ldots (21)$$

This gives the path of the accelerated particles.

Particles can be accelerated in the supernova shock waves associated with the supernova and gamma ray bursts. In general the maximum energy to which the particles can be accelerated in these shock waves is given by:[13, 14]

$$W_{max} \approx (constant) Q B_0 E^{1/3} V_{S0}^{1/3} N_O^{-1/3} \mu^{-1/3} \qquad \ldots (22)$$

Where $Q = Ze$

For a typical type II supernova, the explosion energy $E \approx 10^{51} erg$, $B_0 \approx 10^{-6} G$, $N_0 \approx 1 cm^{-3}$, $\mu = 2$, $V_{S0} \approx 10^{10} cm/s$. Using these in the first equation, we have,[13]

$W_{max} \approx 10^{15} eV$

For example in the case of calcium[13], that is $Z = 20$, we have $W_{max} \approx 10^{16} eV$

Calcium nuclei that are over abundant (by a factor of 10) in some supernova ejecta, can be accelerated to a maximal energy of $\sim 10^{16} eV$, thus accounting for both high energy and calcium anomaly seen in observed high energy cosmic rays.

For a varying magnetic field, say, $B \propto R_S^{-n/2} \propto \left(\dfrac{E}{\rho_0}\right)^{-1/2(5-n)} t^{-1/(5-n)}$, the particle energy takes the form:[13, 14, 15]

$$W = \int V_{S0}^2 \left(\dfrac{E}{\rho_0}\right)^{-1/2(5-n)} t^{-1/(5-n)} t^{2(n-3)/(5-n)} dt \qquad \ldots (23)$$

Substituting for $V_S$ from

$$\int_{p_0}^{p_f} V dp \leq 0.75 Q \int_0^t B_0 V_S^2 dt \qquad \ldots (24)$$

and integrating from $t = t_0$ ($t_0$ is the time when supernova enters the Sedov phase) to some time $t$, we get the maximal particle energy at $t$ as:

$$W < W_0 + 3.75 Q B_0 V_{S0}^2 t_0 \left[ 1 - \left( \frac{t}{t_0} \right)^{(n-1)/(5-n)} \right] \qquad \ldots (25)$$

$W_0$ is the energy gained in pre-Sedov phase.

At $t \gg t_0$, $W$ tends to a maximal limit.

However for a shock wave associated with gamma ray burst, the energy is of the same order, but the magnetic field is much higher, of the order of $B_0 \approx 10^6 G$. However, the number density is also much higher of the order of $\sim 10^{20}/cc$.

This gives the maximum energy to which the particles are accelerated in shock waves associated with gamma ray bursts as

$W_{max} \approx 10^{19} - 10^{20} eV$

For higher Z, this would be higher.

This shows that gamma ray bursts can accelerate protons to the highest energies seen in cosmic rays.[16, 17]

## 5. Synchrotron and Inverse Compton Radiation from the Accelerated Particles

The particles that are accelerated through high magnetic fields in the jets will emit synchrotron radiations. The accretion disk surrounding a black hole produces a thermal spectrum. The lower energy photons produced from this spectrum are scattered to higher energies by relativistic electrons in the surrounding corona, resulting in the inverse Compton radiation. For the particles in the jets, these effects are important.

The synchrotron frequency is given by:

$$\omega_{Syn} = \gamma^2 \omega_B \qquad \ldots (26)$$

Where the frequency of gyration, $\omega_B = \dfrac{eB}{2\pi n_e c} \approx 10^{11}\,Hz$ and the Lorentz factor for the electron $\gamma = \dfrac{E}{m_0 c^2} \approx 10^{10}$. This gives the synchrotron frequency as:

$$\omega_{Syn} \approx 10^{31}\,Hz \qquad \ldots (27a)$$

And the corresponding energy of the electrons is given by:

$$E = h\omega_{Syn} \approx 10^4\,ergs \qquad \ldots (27b)$$

The total power for an isotropic distribution of synchrotron radiation is given by:[18]

$$\dot{E} = \dfrac{4}{3}\sigma_T c\beta^2 \left(\dfrac{B^2}{8\pi}\right)\Gamma^2 \qquad \ldots (28)$$

Where, $\sigma_T$ is the Thompson cross section.

The radiation power is given by:

$$\dfrac{P}{4\pi r_i^2 \Gamma^2 c} \qquad \ldots (29)$$

Where the power $P \approx 10^{51}\,ergs/s$, as given by equation (7) and $r_i = \Gamma^2 c\Delta t$. If $\varepsilon$ (say 20%) is the fraction of this power converted to magnetic field, then[13, 19]

$$\dfrac{B^2}{8\pi} = \varepsilon \dfrac{P}{4\pi \Delta t^2 c^3 \Gamma^6} \qquad \ldots (30)$$

The magnetic field works out to be of the order of $10^4\,G$, where the duration of the event $\Delta t \sim 1s$ and gamma factor ~300.

Using these data in equation (24), the total power of synchrotron radiation emitted by each electron works out to be:

$$\dot{E} \sim 10^{-3}\,ergs/s \qquad \ldots (31)$$

The power radiated by the particles during the inverse Compton radiation is given by:

$$\dot{E} = \sigma_T U_\gamma \Gamma^2 \beta c \qquad \ldots (32)$$

Where the energy density for a power $P \approx 10^{51}\,ergs/s$, is given by:

$$U_\gamma = \dfrac{P}{4\pi r^2 c} \approx 10^{19}\,ergs/cc \qquad \ldots (33)$$

Using this in equation (28), the power radiated during the inverse Compton radiation by an electron works out to be:

$$\dot{E} \sim 10^9 \, ergs/s \qquad \ldots (34)$$

For both synchrotron and IC power to be comparable, that is

$$U_\gamma = \frac{B^2}{8\pi} \qquad \ldots (35)$$

The magnetic field should be of the order of $10^{10} G$. Form flux conservation, this corresponds to a distance given by:

$$R = \frac{B_0}{B} R_0 = 10^9 \, cm \qquad \ldots (36)$$

Where, $B_0 = 10^{16} G$ is the surface magnetic field at $R_0 = 10^6 \, cm$.

The radiation power[20] is given by equation (25) as $\frac{P}{4\pi r_i^2 \Gamma c}$. If $m_0 c^2$ is the energy associated with each of the particles, then the number density is given by:

$$n = \frac{P}{4\pi r_i^2 \Gamma c (m_0 c^2)} \approx 10^{11} /cc \qquad \ldots (37)$$

The total energy released due to the IC radiation is given by:

$$E = \dot{E} n \left( \frac{4}{3} \pi R^3 \right) \approx 10^{51} \, ergs \qquad \ldots (38)$$

At a distance of $10^9 \, cm$, both, synchrotron and IC radiation energy released becomes equal, as given by equation (36).

The energy radiated by the particles due to IC or synchrotron radiation is given by equation (32) as $\dot{E} = \sigma_T U \Gamma^2 \beta c$, where $U$ is the corresponding energy density. Integrating and for the limit that the energy reduce by half, we get the half-life as:[18]

$$t_{1/2} = \frac{3 m_e^4 c^7}{2 e^4 U E} \qquad \ldots (39)$$

Where, the energy of each particle is given by equation (27b), i.e, $E \approx 10^4 \, ergs$

In the case of IC radiation, the energy density $U_\gamma \approx 10^{19} \, ergs/cc$, and that for the synchrotron radiation is given by $U_{Syn} = \dfrac{B^2}{8\pi} \approx 10^7 \, ergs/cc$.

Therefore, the half-life for an electron becomes:

$$\left(t_{1/2}\right)_{IC} \approx 10^{-21} \, s$$
$$\left(t_{1/2}\right)_{Syn} \approx 10^{-9} \, s \qquad \ldots (40)$$

As we see, the half-lives for an electron are extremely small and they will rapidly lose their energies.

In the case of protons[18], the synchrotron frequency is given by equation (26) as, $\omega_{Syn} = \gamma^2 \omega_B$, where the frequency of gyration, $\omega_B = \dfrac{eB}{2\pi m_p c} \approx 10^8 \, Hz$ and the Lorentz factor for the proton $\gamma = \dfrac{E}{m_p c^2} \approx 10^7$.

This gives the synchrotron frequency as:

$$\omega_{Syn} \approx 5 \times 10^{21} \, Hz \qquad \ldots (41a)$$

And the corresponding energy is given by:

$$E = h\omega_{Syn} \approx 10^{-5} \, ergs \qquad \ldots (41b)$$

The half-life for the proton then becomes:

$$\left(t_{1/2}\right)_{IC} \approx 3s$$
$$\left(t_{1/2}\right)_{Syn} \approx 10^{12} \, s \qquad \ldots (42)$$

Due to the fact that the half-life scales as $m^4$, protons and heavier ions, such as Fe, Ca, etc., will retain their energies for a longer time. So we expect high energy protons from these sources and they could be a possible source of high energy cosmic rays.

## 6. Synchrotron and Inverse Compton Radiation from galactic jets

The same arguments from section 5 will apply to the galactic jets. The difference being that the magnetic field is much smaller (of the order of $10^{-6} G$) but spanning over a larger scale of about $10^{23} cm$.

The potential experienced by the electrons is given by $V = \int E_0 dx$ and over the galactic scales. In the galaxy, the magnetic field remains more or less uniform over the scale of $\sim 10^{23} cm$. So the field is of the order of $10^{-6} - 10^{-7} G$.

So over the galactic scale the potential works out to be of the order of $V \approx 10^{16} V$.

The energy gained by each electron is therefore $E_0 \approx 10^{16} eV$. The acceleration of the electrons is given by:

$$a = \frac{eE_0}{m_0} \approx 10^{33} cm/s^2 \qquad \ldots (43)$$

The synchrotron frequency is given by equation (26) as, $\omega_{Syn} = \gamma^2 \omega_B$. The frequency of gyration, $\omega_B = \frac{eB}{2\pi m_e c} \approx 1 Hz$ and the Lorentz factor for the electron $\gamma = \frac{E}{m_0 c^2} \approx 10^{10}$.

This gives the synchrotron frequency as:

$$\omega_{Syn} \approx 10^{20} Hz \qquad \ldots (44a)$$

And the corresponding energy of the electrons is given by:

$$E = h\omega_{Syn} \approx 10^{-6} ergs \qquad \ldots (44b)$$

The total power for an isotropic distribution of synchrotron radiation is given by equation (28):

$$\dot{E} = \frac{4}{3}\sigma_T c\beta^2 \left(\frac{B^2}{8\pi}\right)\Gamma^2 \approx 10^{-6} ergs/s \qquad \ldots (45)$$

The number density is given by equation (37) as:

$$n = \frac{P}{4\pi r_i^2 \Gamma c(m_0 c^2)} \approx 10^{-6}/cc \qquad \ldots (46)$$

The total energy released due to the synchrotron then becomes:

$$E = \dot{E}n\left(\frac{4}{3}\pi R^3\right) \approx 10^{63} ergs \qquad \ldots (47)$$

The half life for the synchrotron radiation is given by equation (39)

$$t_{1/2} = \frac{3m_e^4 c^7}{2e^4 UE} \approx 10^{18} s \qquad \ldots (48)$$

In the case of the proton, the synchrotron frequency is given by:[18]

$$\omega_{Syn} = \gamma^2 \omega_B \approx 10^{11} Hz \qquad \ldots (49a)$$

Where, the frequency of gyration, $\omega_B = \frac{eB}{2\pi m_e c} \approx 10^{-3} Hz$ and the Lorentz factor

$\gamma = \frac{E}{m_0 c^2} \approx 10^7$.

And the corresponding energy is given by:

$$E = h\omega_{Syn} \approx 10^{-16} ergs \qquad \ldots (49b)$$

The volume of the galactic jet is given by

$$V = \frac{\pi r^2 l}{3} = \frac{\pi l^3 \tan^2 5°}{3} \qquad \ldots (50)$$

Where, $l$ is the length of the jet and the opening angle of the jet is $\sim 5°$.

For a jet of mega parsec length (see section 2), the volume $V \approx 10^{68} cc$

From equation (13) we have that the particles in the jet are accelerated to energies of $10^{16} eV$. Assuming equipartition of energy, the number density of particles accelerated to this energy, $n = 10^{-16} /cc$

The power associated with each electron is given by equation (28). For an energy density of $\frac{B^2}{8\pi} \sim 10^{-14} ergs/cc$, this works out to be

$$\dot{E} = \frac{4}{3} \sigma_T c \beta^2 \left(\frac{B^2}{8\pi}\right) \Gamma^2 \sim 10^{-9} ergs/s/\text{electron} \qquad \ldots (51)$$

The power associated with the total number of electrons $n$, is

$$\dot{E}n \approx 10^{-25} ergs/s/cc \qquad \ldots (52)$$

And over the entire volume, this works out to be

$$\left(\dot{E}\right)_{Vol} = n\dot{E}V \approx 10^{44} ergs/s \qquad \ldots (53)$$

## 7. Concluding Remarks

In this paper we have discussed the dynamics of jets powered by black hole rotation and magnetic fields. The scaling of the jet length with black hole mass and their life times are consistent with recent observations. We have also arrived at the trajectories of accelerated particles in the jets and discussed the possibility of these being source of high energy cosmic rays.